\documentclass[final,5p,times,twocolumn]{elsarticle}
\usepackage{lineno}
\usepackage{soul}
\biboptions{sort&compress}

\journal{Physics Letters B}

\usepackage{amsmath}
\usepackage{caption}
\usepackage{subcaption}
 
\usepackage{bm}
\usepackage{color}
\usepackage{graphicx,epsfig}
\usepackage{amssymb}
\usepackage{multirow}

\usepackage{dcolumn}     % Align table columns on decimal point
\newcolumntype{.}{D{.}{.}{-1}}

\newcommand{\LO}{{\text{LO}}}
\newcommand{\NLO}{{\text{NLO}}}
\newcommand{\NNLO}{{\text{N}^{2}\text{LO}}}
\newcommand{\NNNLO}{{\text{N}^{3}\text{LO}}}
\newcommand{\NNNNLO}{{\text{N}^{4}\text{LO}}}
\newcommand{\NkLO}{{\text{N}^{k}\text{LO}}}

\newcommand{\myvec}[1]{{\boldsymbol {#1}  }}

\newcommand{\EKM}{{\text{N}^{k}\text{LO}_{EKM}}}
\newcommand{\EMN}{{\text{N}^{k}\text{LO}_{EMN}}}

\newcommand\Tstrut{\rule{0pt}{5.6ex}}         % = `top' strut
\newcommand\Bstrut{\rule[-0.9ex]{0pt}{0pt}}   % = `bottom' strut

\newcommand{\LOemnAvg}{{-1.599}}
\newcommand{\LOemnMin}{{-0.097}}
\newcommand{\LOemnMax}{{+0.095}}
\newcommand{\LOemnMinTot}{{-0.099}}
\newcommand{\LOemnMaxTot}{{+0.097}}

\newcommand{\NLOemnAvg}{{-1.710}}
\newcommand{\NLOemnMin}{{-0.029}}
\newcommand{\NLOemnMax}{{+0.029}}
\newcommand{\NLOemnMinTot}{{-0.035}}
\newcommand{\NLOemnMaxTot}{{+0.035}}

\newcommand{\NNLOemnAvg}{{-1.710}}
\newcommand{\NNLOemnMin}{{-0.009}}
\newcommand{\NNLOemnMax}{{+0.008}}
\newcommand{\NNLOemnMinTot}{{-0.022}}
\newcommand{\NNLOemnMaxTot}{{+0.022}}

\newcommand{\NNNLOemnAvg}{{-1.712}}
\newcommand{\NNNLOemnMin}{{-0.005}}
\newcommand{\NNNLOemnMax}{{+0.006}}
\newcommand{\NNNLOemnMinTot}{{-0.021}}
\newcommand{\NNNLOemnMaxTot}{{+0.021}}

\newcommand{\NNNNLOemnAvg}{{-1.712}}
\newcommand{\NNNNLOemnMin}{{-0.006}}
\newcommand{\NNNNLOemnMax}{{+0.006}}
\newcommand{\NNNNLOemnMinTot}{{-0.021}}
\newcommand{\NNNNLOemnMaxTot}{{+0.021}}
\newcommand{\LOekmAvg}{{-1.767}}
\newcommand{\LOekmMin}{{-0.17}}
\newcommand{\LOekmMax}{{+0.18}}
\newcommand{\LOekmMinTot}{{-0.17}}
\newcommand{\LOekmMaxTot}{{+0.18}}

\newcommand{\NLOekmAvg}{{-1.718}}
\newcommand{\NLOekmMin}{{-0.034}}
\newcommand{\NLOekmMax}{{+0.025}}
\newcommand{\NLOekmMinTot}{{-0.040}}
\newcommand{\NLOekmMaxTot}{{+0.032}}

\newcommand{\NNLOekmAvg}{{-1.705}}
\newcommand{\NNLOekmMin}{{-0.010}}
\newcommand{\NNLOekmMax}{{+0.008}}
\newcommand{\NNLOekmMinTot}{{-0.023}}
\newcommand{\NNLOekmMaxTot}{{+0.022}}

\newcommand{\NNNLOekmAvg}{{-1.719}}
\newcommand{\NNNLOekmMin}{{-0.012}}
\newcommand{\NNNLOekmMax}{{+0.009}}
\newcommand{\NNNLOekmMinTot}{{-0.024}}
\newcommand{\NNNLOekmMaxTot}{{+0.022}}

\newcommand{\NNNNLOekmAvg}{{-1.718}}
\newcommand{\NNNNLOekmMin}{{-0.009}}
\newcommand{\NNNNLOekmMax}{{+0.008}}
\newcommand{\NNNNLOekmMinTot}{{-0.022}}
\newcommand{\NNNNLOekmMaxTot}{{+0.022}}

\newcommand{\LOsimAvg}{{-1.616}}
\newcommand{\LOsimMin}{{-0.11}}
\newcommand{\LOsimMax}{{+0.11}}
\newcommand{\LOsimMinTot}{{-0.11}}
\newcommand{\LOsimMaxTot}{{+0.11}}

\newcommand{\NLOsimAvg}{{-1.724}}
\newcommand{\NLOsimMin}{{-0.032}}
\newcommand{\NLOsimMax}{{+0.032}}
\newcommand{\NLOsimMinTot}{{-0.038}}
\newcommand{\NLOsimMaxTot}{{+0.038}}

\newcommand{\NNLOsimAvg}{{-1.721}}
\newcommand{\NNLOsimMin}{{-0.011}}
\newcommand{\NNLOsimMax}{{+0.011}}
\newcommand{\NNLOsimMinTot}{{-0.023}}
\newcommand{\NNLOsimMaxTot}{{+0.023}}

\usepackage{graphicx}
\usepackage{capt-of}
\usepackage{amsmath}

\usepackage{amssymb}
\usepackage{relsize}
\usepackage[colorlinks,breaklinks]{hyperref}
\usepackage{xcolor}

\usepackage{multirow}

\usepackage{comment}

\begin{document}

\begin{frontmatter}

\title{The deuteron-radius puzzle is alive: a new analysis of nuclear structure uncertainties}

\author[Mainz,UBC,Triumf]{O.~J.~Hernandez}
\ead{javierh@phas.ubc.ca}

\author[Chalmers]{A.~Ekstr\"om}
\ead{andreas.ekstrom@chalmers.se}

\author[Triumf]{N.~Nevo Dinur}
\ead{nnevodinur@triumf.ca}

\author[NormalU]{C.~Ji}
\ead{jichen@mail.ccnu.edu.cn}

\author[Mainz,Triumf,Manitoba]{S.~Bacca}
\ead{s.bacca@uni-mainz.de}

\author[Hebrew]{N.~Barnea}
\ead{nir@phys.huji.ac.il}

\address[Mainz]{Institut f\"ur Kernphysik and PRISMA Cluster of Excellence, Johannes Gutenberg-Universit\"at Mainz, 55128 Mainz, Germany}
\address[UBC]{Department of Physics and Astronomy, University of British Columbia, Vancouver, BC, V6T 1Z4, Canada}
\address[Triumf]{TRIUMF, 4004 Wesbrook Mall, Vancouver, BC V6T 2A3, Canada}
\address[Chalmers]{Department of Physics, Chalmers University of Technology, SE-412 96 Gothenburg, Sweden}
\address[NormalU]{Key Laboratory of Quark and Lepton Physics (MOE) and Institute of Particle Physics,  Central China Normal University, Wuhan 430079,China}
\address[Manitoba]{Department of Physics and Astronomy, University of Manitoba, Winnipeg, MB, R3T 2N2, Canada}
\address[Hebrew]{Racah Institute of Physics, The Hebrew University, Jerusalem 9190401, Israel}

\begin{abstract}
To shed light on the deuteron radius puzzle we analyze the
theoretical uncertainties of the nuclear structure corrections to the
Lamb shift in muonic deuterium. We find that the discrepancy between
the calculated two-photon exchange correction and the corresponding
experimentally inferred value by Pohl {\it et al.}~\cite{Pohl_2016}
remain. The present result is consistent with our previous estimate,
although the discrepancy is reduced from 2.6~$\sigma$ to about
2~$\sigma$. The error analysis includes statistic as well as
systematic uncertainties stemming from the use of
nucleon-nucleon interactions derived from chiral effective field
theory at various orders. We therefore conclude that   nuclear theory
uncertainty is more likely not the source of the discrepancy.
\end{abstract}
\begin{keyword}
\texttt{elsarticle.cls}\sep \LaTeX\sep Elsevier \sep template
\MSC[2010] 00-01\sep  99-00
\end{keyword}

\end{frontmatter}

%\linenumbers

\section{Introduction}

The charge radius of the deuteron ($d$), the simplest nucleus consisting
of one proton and one neutron, was recently determined to be
$r_d$ = 2.12562(78) fm \cite{Pohl_2016} 
using several
Lamb shift (LS) transitions in muonic deuterium ($\mu-d$). This result
provides three times the precision compared with previous
measurements. Furthermore, the $\mu-d$ value is 7.5~$\sigma$ or
5.6~$\sigma$ 
smaller than the world averaged
CODATA-2010~\cite{CODATA_2010} or CODATA-2014~\cite{CODATA_2014}
values, respectively, and  3.5~$\sigma$ smaller than the result from
ordinary deuterium  spectroscopy~\cite{Pohl_2017}. 
One can also combine the measured radius squared
difference $r_d^2 - r_p^2$
obtained from isotope shift experiments on ordinary hydrogen and
deuterium~\cite{Parthey2010} with the absolute determination of the
proton radius from muonic hydrogen
experiments~\cite{Pohl_2010,Antognini13} (dubbed as ``$\mu p+$iso'') to obtain $r_d = 2.12771(22)$ fm, which is much closer to the $\mu-d$ result, but still differs from it by 2.6 $\sigma$ (see Ref.~\cite{Pohl_2016} for details).
Altogether,
these significant discrepancies have been coined ``the deuteron radius
puzzle''.

Unlike with the proton-radius puzzle~\cite{Pohl_2010}, 
$r_d$ from $\mu-d$ Lamb shift measurements
is consistent with the electron-deuteron ($e-d$) scattering
data due to the
large uncertainty in the scattering experiments. 
Ongoing efforts to improve the precision in electron scattering
will provide further information~\cite{Mainz_exp}. However, these
discrepancies, compounded with the 7~$\sigma$ (5.6~$\sigma$)
discrepancy between the CODATA-2010 (CODATA-2014) and the muonic
hydrogen proton radius~\cite{Pohl_2010,Antognini13}, highlight the
need to pinpoint the source of the differences. While
the very recent $2S-4P$ spectroscopy
on ordinary hydrogen supports the  small proton radius~\cite{Beyer_Science_2017}, the
conundrum of the proton and deuteron radius puzzles is not yet fully
solved and further experimental and theoretical investigations are
clearly required.

\begin{figure}[h]
  \centering
    \includegraphics[width=0.14\textwidth]{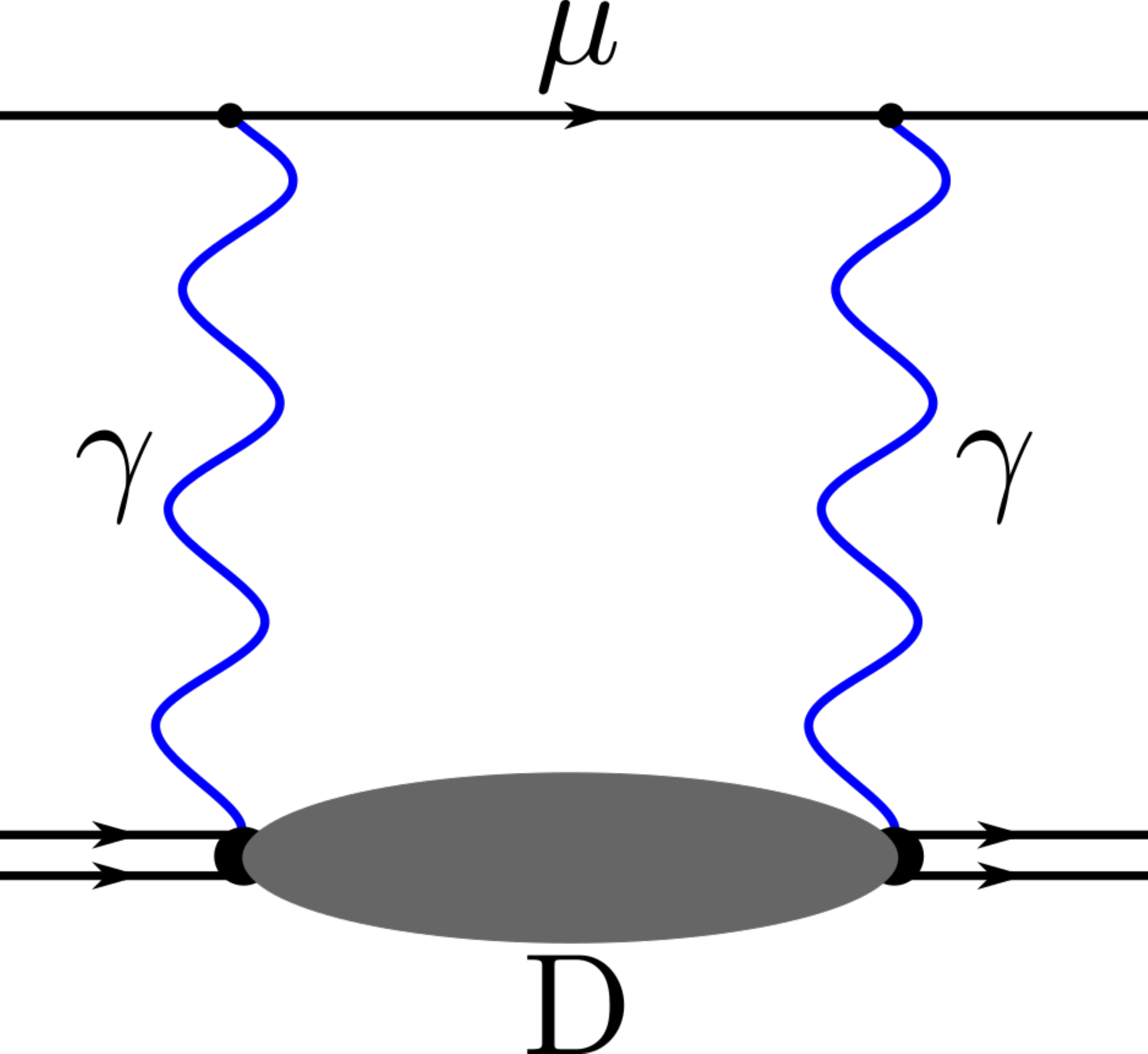}
      \caption{Feynman diagram of the two-photon exchange between the muon and the deuteron.}
\label{fig:tpe}
\end{figure}

The deuteron charge radius $r_d$ is extracted
from the LS measurement through
\begin{equation}
\label{eq:Delta_LS and Delta_TPE relations}
\Delta E_{\text{LS}} = \delta_{\text{QED}}+\delta_{\text{TPE}}+\frac{m_r \alpha^4}{12}{r_d^2},
\end{equation}
which is valid in an $\alpha$ expansion up to 5$^{th}$ order, where
$\alpha$ is the fine structure constant. The term $m_r$ in
Eq.~(\ref{eq:Delta_LS and Delta_TPE relations}) is the reduced mass of
the $\mu-d$ system. The LS energy difference, $\Delta E_{\text{LS}}$,
is directly measured through pulsed laser spectroscopy experiments
described in detail in \cite{Pohl_2016,Pohl_2010,Antognini13,Antognini_2013}. 
The quantum
electrodynamic (QED) corrections $\delta_{\text{QED}}$ are obtained
from highly accurate theoretical calculations~\cite{Borie:2012zz,Krutov:2011ch}.  In
the extraction of $r_d$ from LS measurements
the main source of uncertainty is due to nuclear structure corrections
coming from a two-photon exchange (TPE) diagram, $\delta_{\rm
  TPE}$ depicted in Fig.~\ref{fig:tpe}. Because the latter is obtained from theoretical computations
it is of paramount importance that all theoretical uncertainty
contributions are thoroughly investigated.

Several groups have calculated $\delta_{\rm TPE}$ with different
methods~\cite{Pachucki_2011,Friar_2013,Carlson_2014,Hernandez_2014,Pachucki_2015,Hernandez_2016}. The
two most recent and most precise computations,
Ref.~\cite{Pachucki_2015} and Ref.~\cite{Hernandez_2016}, are
consistent within $0.6\%$. All theoretical calculations have been
summarized by Krauth {\it et al.}~\cite{Krauth_2016} which resulted in
a recommended value of $\delta_{\text{TPE}}$=-1.7096(200) meV. This
value was also used by Pohl {\it et al.}~\cite{Pohl_2016} to extract
$r_d$ using Eq.~(\ref{eq:Delta_LS and Delta_TPE
  relations}).

On the other hand, measuring $\Delta E_{\rm LS}$ and knowing $\delta_{\rm QED}$,
Eq.~(\ref{eq:Delta_LS and Delta_TPE relations}) enables the
extraction of $\delta_{\rm TPE}$ from an experimentally determined
radius. Using $r_d$ from ``$\mu p+$iso'' leads to an experimental value  $\delta_{\text{TPE}}$=-1.7638(68) meV~\cite{Pohl_2016}, which differs from the theoretical one by 2.6 $\sigma$. This disagreement motivates a reassessment of the theoretical calculation, and in particular of its assigned uncertainties.

Finally, from the radii of light
nuclei, such as hydrogen and deuterium, it is possible to determine
the Rydberg constant $R_{\infty}$ and consequently the radius puzzle
can be turned into a ``Rydberg constant puzzle''. In
Ref.~\cite{Pohl_2017} two values of $R_{\infty}$ were calculated using
the muonic hydrogen and muonic deuterium charge radii separately and
the results were found to disagree by 2.2 $\sigma$. This difference
was attributed to the $\delta_{\rm TPE}$ contribution used to extract
 $r_d$ from the Lamb shift.

The purpose of this letter is to revisit our calculations of the
nuclear structure corrections in $\mu-d$ and exploit chiral effective
field theory and statistical regression analysis to systematically
improve the theoretical uncertainty estimation in $\delta_{\rm TPE}$ and shed light
on the deuteron radius puzzles.

State-of-the-art calculations of $\delta_{\text{TPE}}$ in
Refs.~\cite{Hernandez_2014,Hernandez_2016} as well as in this work,
employ nucleon-nucleon (NN) potentials derived from a low-energy
expansion of quantum chromodynamics called chiral effective field
theory (chiral EFT). Within this approach, which also constitutes the
modern paradigm of analyzing the nuclear interaction, the nuclear
potential is built from a sum of  pion-exchange contributions 
and nucleon contact terms, see,
e.g., Refs.~\cite{epelbaum2009,machleidt2011}. Power counting enables
to determine the importance of individual terms in the low-energy
expansion and thereby also facilitates a meaningful truncation of higher-order diagrams that build the potential. All
potentials in this work employ  Weinberg's dimensional power counting schemes~\cite{Weinberg:1990rz,Weinberg:1991um},
whereby the order $\nu \geq 0$ to which a diagram belongs is
proportional to $Q^{\nu}$, where
\begin{equation}
\label{eq: ChiEFT expansion parameter}
Q = \text{max}\left \{ \frac{p}{\Lambda_b}, \frac{m_{\pi}}{\Lambda_b} \right \}\,
\end{equation}
and $p$ is a small external momentum, $\Lambda_b$ is the chiral
symmetry breaking scale of about the rho meson mass, and $m_{\pi}$ is
the pion mass. Given a power counting, contributions with a low power
of $\nu$ are more important than terms at higher powers. Starting from
the leading order (LO), i.e., $\nu=0$, higher orders will be denoted
as next-to-leading order (NLO), i.e., $\nu=2$, next-to-next-to-leading
order N$^2$LO, i.e., $\nu=3$, and so on. It is worth noticing that, in
chiral EFT, the contributions with $\nu=1$ vanish due to time-reversal
and parity. At each order $\nu$ of the chiral EFT potential, there
will be a finite set of parameters, known as low energy constants
(LECs), that determine the strength of various pion-nucleon and
multi-nucleon operators. The LECs are not provided by the theory
itself but can be obtained from fitting to selected experimental data,
such as NN and $\pi$N scattering cross sections, and other few-body
ground state observables, such as radii and binding
energies. Different fitting procedures exist, and we will explore a
variety of them as a way to probe both statistical and systematic uncertainties.

To avoid infinities upon iteration in the Lippmann-Schwinger equation
all chiral potentials are regulated by exponentially suppressing
contributions with momenta $p$ greater than a chosen cutoff value
$\Lambda$, see
e.g. Refs~\cite{epelbaum2009,machleidt2011}. Non-perturbative ab
initio calculations using momentum-space chiral EFT often employ NN
interactions with $\Lambda \approx 400-600$ MeV.

Chiral EFT and effective field theories in general, unlike
phenomenological models, furnish a systematic, i.e, order-by-order,
description of low-energy processes at a chosen level of
resolution. In this work, it provides us with an opportunity to
estimate the uncertainty of $\delta_{\text{TPE}}$ truncated up to
different chiral orders. In our previous
work~\cite{Hernandez_2014,Hernandez_2016}, we probed the theoretical
uncertainty stemming from the nuclear physics models by cutoff
variation, i.e., varying $\Lambda$. Strictly speaking, this
prescription to estimate the chiral EFT uncertainty also requires 
the excluded $\nu$th-order chiral contributions  to be
proportional to $1/\Lambda^{\nu+1}$ when $\Lambda$ is approaching the breakdown scale $\Lambda_b$: 
a property that
hinges on order-by-order renormalizability of the canonical chiral EFT
formulation, which is not yet established. Also, cutoff-variation tend
to either underestimate or overestimate the chiral EFT systematic
uncertainty with respect to the variation range~\cite{Epelbaum_2015_01,Epelbaum_2015_02}. To this end, and
to be as conservative as possible, we will augment the procedure of
cutoff-variation by implementing the chiral EFT truncation-error to obtain solid systematic uncertainty estimates.

%and our final result was in agreement at the 0.6$\%$ level and within
%uncertainties with another very accurate calculation by Pachucki and
%Wienczek~\cite{Pachucki_2015}.

Any rigorous estimate of the theoretical uncertainty must also
consider the effects of the statistical uncertainties of the
LECs due to experimental uncertainties in the pool of fitted 
data. For example, in Ref.~\cite{Acharya_2016} it was found that a
rigorous statistical analysis lead to a four-fold increase in the
uncertainty estimates of the proton-proton fusion $S$-factor as
compared to previous work which only probed the systematic uncertainty
of the nuclear model by limited cutoff variations. Motivated by the
possibility that the uncertainties were underestimated, we
rigorously probe the statistical and systematic uncertainties in the
nuclear structure corrections in the Lamb shift of $\mu-d$, by
propagating the uncertainties of the LECs appearing in the NN
potentials up to N$^2$LO \cite{Ekstrom_2013,Carlsson_2016}.

Details on the observables associated with the LS in $\mu-d$ are
explained in Section~\ref{tpe} and results of the statistical analysis
will be shown in Section~\ref{stat}. In addition, we improve our
estimates of the systematic uncertainty associated with the chiral EFT
expansion by carrying out our calculations up to fifth-order in chiral
EFT, namely N$^4$LO. We then use the method detailed in
Refs.~\cite{Epelbaum_2015_01,Epelbaum_2015_02,Furnstahl_2015} to estimate
the systematic uncertainty associated with the chiral truncation at
each order. Results will be shown in Section~\ref{syst}. Finally, we
will examine and combine all the relevant sources of uncertainty in Section~\ref{results}, before drawing
conclusions in Section~\ref{conclusion}.

\section{Two-photon exchange contributions}
\label{tpe}

For the calculation of $\delta_{\rm TPE}$ we separate terms that depend on the few-nucleon dynamics, denoted with $A$, from terms that exclusively depend on properties of the single-nucleon,
denoted with $N$, as 
\begin{equation}
  \delta_{\text{TPE}} =\delta^A_{\rm TPE}+ \delta^N_{\rm TPE}.
\end{equation}
The first contribution, which is the focus of this paper, can be
written as
\begin{equation}
  \label{TPE_A}
  \delta^A_{\rm TPE} =  \delta^{(0)}+\delta^{(1)}+\delta^{(2)}+ \delta^{A}_{Zem}+\delta^{(1)}_{NS}+\delta^{(2)}_{NS}\,.
\end{equation}
Here $ \delta^{(0)}$ is the leading, $\delta^{(1)}$ the subleading and $\delta^{(2)}$ the sub-subleading term 
 in the context of an $\eta$-expansion. The small parameter  $\eta $ is $\sqrt{m_r/m_d}$ with $m_d$ being
 the deuteron mass.  The term $\delta^{A}_{Zem}$ is the third Zemach moment.
 These first four terms are  calculated in the point-nucleon limit, while nucleon size ($NS$) corrections
are added with the additional terms $\delta^{(1)}_{NS}$ and $\delta^{(2)}_{NS}$.
The formulas  of these corrections  are detailed in
Refs.~\cite{Ji_2013,Hernandez_2014,Hernandez_2016,Nevo_Dinur_2016} and not repeated here.

The single-nucleon terms, which we take from the literature, can be decomposed into
  \begin{equation}
      \label{TPE_N}
\delta^N_{\rm TPE}=\delta_{Zem}^{N}+\delta_{pol}^N+\delta_{sub}^N\,.
\end{equation}
  The nucleon Zemach moment and polarizability terms are estimated from data and taken to be $\delta_{Zem}^{N}=-0.030(2)$ meV~\footnote{This value is estimated by rescaling the $\mu H$ value adopted by \cite{Krauth_2016} according to the scaling factor described in Refs.~\cite{Hernandez_2016,Nevo_Dinur_2016}.} and $\delta_{pol}^N= -0.028(2)$ meV~\cite{Carlson_2014}. The subtraction term $\delta_{sub}^N$ instead is not well constrained by experimental data. Its contribution in $\mu$H was calculated by Birse and McGovern using chiral perturbation theory to be 0.0042(10) meV~\cite{Birse:2012eb}. With the operator product expansion approach, Hill and Paz recalculated the TPE in $\mu$H~\cite{Hill:2016bjv}, obtaining a central value that agrees with Ref.~\cite{Birse:2012eb} but has a much larger uncertainty (see also Refs.~\cite{Birse:2017czd,Hill:2017rlj}). For the nucleonic subtraction in $\mu$D, we adopt the strategy of Ref.~\cite{Krauth_2016} assuming that proton and neutron subtraction terms are of the same size and assigning a 100\% uncertainty. This yields $\delta_{sub}^N=$ 0.0098(98) meV~\cite{Krauth_2016}. If we enlarge the uncertainty to 200\% to be comparable with the finding in Ref.~\cite{Hill:2016bjv}, we have $\delta_{sub}^N=$ 0.0098(196) meV.

The calculations for the deuteron are based on a harmonic oscillator
expansion of the wave function and a discretize approach to the sum
rules to compute the terms in Eq.~(\ref{TPE_A}), see Refs.~\cite{Hernandez_2014,Hernandez_2016}. We
have previously shown that this approach is reliable and agrees very
well with other calculations, such as from
Pachucki~\cite{Pachucki_2011} and Arenh\"{o}vel~\cite{Arenhovel_book}.

\section{Statistical uncertainty estimates}
\label{stat}

To probe the uncertainties in $\delta_{\rm TPE}$ due to statistical
uncertainties in the LECs -- originating from the uncertainties in the
pool of fitted data -- we use the $\NkLO_{sim}$ potentials from
Ref.~\cite{Carlsson_2016}, with $k$ from 0 to 2\footnote{Note that $k$
  and $\nu$ are not exactly the same, even though there is a
  one-to-one correspondence between them; $k=0,1,2,3,4$ corresponds to
  $\nu=0,2,3,4,5$.}. The presently available $sim$ potentials employ
seven different cutoff values $\Lambda=450,475,\ldots,575,600$ MeV,
and for each cutoff a potential was simultaneously fit to six
increasingly larger energy-ranges of the SM99 world database of NN
scattering cross sections, along with $\pi$N scattering cross
sections, and ground state properties\footnote{Radius, energy, and for
  $^{2}$H also the quadrupole moment.} of the $^{2,3}$H and $^3$He
systems. The various subsets of the NN scattering data are delimited
by the maximum kinetic energy in the laboratory frame of reference;
$T^{\rm max}_{\rm Lab}=125,158,191,224,257,290$ MeV. The statistical
covariance matrix of the LECs for each $sim$ interaction was
determined numerically to machine precision using forward-mode
automatic differentiation. The covariance matrices make it possible to
propagate the statistical uncertainties of the LECs to, e.g.,
$\delta_{\rm TPE}$, and the various $\Lambda$ and $T^{\rm max}_{\rm
  Lab}$ cuts gauge the systematic uncertainties in the fitting
procedure and the cutoff choice.

We compute the covariance matrix of the nuclear structure corrections
relevant to the $\mu-d$ system using the linear approximation. For two
quantities $A$ and $B$, matrix elements of the covariance matrix are
then obtained from
\begin{equation}
\text{Cov}(A,B) = \myvec{J}_{A} \text{Cov}(\alpha) \myvec{J}^{T}_{B}.
\end{equation}
Here, $\myvec{J}_{A}$ is the row vector of partial derivatives of the
quantity $A$, with respect to the set of LECs $\alpha$, and
analogously for $\myvec{J}_B$. The covariance matrix
$\text{Cov}(\alpha)$ of LECs is provided from \cite{Carlsson_2016}. The vector
components $J_{A,i} = \frac{\partial A}{\partial \alpha_{i} }$ are
obtained from a univariate spline fit to ten numerical function
evaluations of $A$ in a small neighborhood of the optimal value of the
LEC $\alpha_{i}$. To benchmark our procedure for calculating the
derivatives in this fashion we compared our results for the deuteron
ground state properties, such as the ground state energy $E_0$, the
root mean-square point-nucleon radius $r$, and the quadrupole
moment $Q_d$, 
with existing computations based on automatic
differentiation algorithms and obtained excellent agreement for all
quantities, i.e. better than 0.005$\%$ relative uncertainty. The
linear correlation coefficient between $A$ and $B$ is given by
\begin{equation}
\rho(A,B)= \frac{\text{Cov}(A,B)}{\sigma_{A,stat} \sigma_{B,stat}},
\end{equation}
where $\sigma_{A,stat} \equiv \sqrt{\text{Cov}(A,A)}$, and similarly for
$\sigma_{B,stat}$, are the statistical uncertainties of $A$ and $B$,
respectively. A value $\rho(A,B)=1$ ($\rho(A,B)=-1$) indicates
fully (anti-) correlated quantities, while $\rho(A,B)=0$ implies that
$A$ and $B$ are uncorrelated.

A regression analysis provides an exact quantification of the
statistical correlations that are present in the chiral EFT
description of the nuclear polarization effects. This allows us to
determine constraints between different observables predicted within
the $\text{N}^{k}\text{LO}_{sim}$ models and serves as a valuable check of
the uncertainty propagation formalism. In this work, we focus on
$\delta^A_{\rm TPE}$ and its components, such as the leading dipole
term $\delta^{(0)}_{D1}$ and the small magnetic term $\delta^{(0)}_{M}$. Since
they are related to the electric dipole and magnetic dipole response, we
also study the electric dipole polarizability $\alpha_E$ and the
magnetic susceptibility $\beta_M$. Detailed expressions can be
found in Refs.~\cite{Hernandez_2014,Hernandez_2016}. We also compute
other observables of interest, such as the ground-state energy $E_0$
and the mean square point-proton distribution radius $r$,
related to $r_d$ by
\begin{equation}
  \label{eq_radius}
r_d^2
= r^2 +r^2_p+r_n^2 + \frac{3}{4m^2} + r^2_{\rm 2BC} \,,
\end{equation}
where $r_{p/n}$ are the proton and neutron radius, respectively, while
$m$ is the proton mass and $r^2_{\rm 2BC}$ is a contribution due to two-body  currents~\cite{Pastore08,Kolling11} and other relativistic corrections beyond the Darwin-Foldy term $\frac{3}{4m^2} $.  Furthermore, we also study other ground-state
observables, such as the electric quadrupole moment $Q_d$ and the
magnetic dipole moment $\mu_d$, along with the $D$-wave probability
$P_D$. The linear correlation coefficients between these
observables are presented in Fig.~\ref{fig:the N2LO systematic and
  statistical uncertainty estimates}.

\begin{figure}[h]
  \centering
    \includegraphics[width=0.48\textwidth]{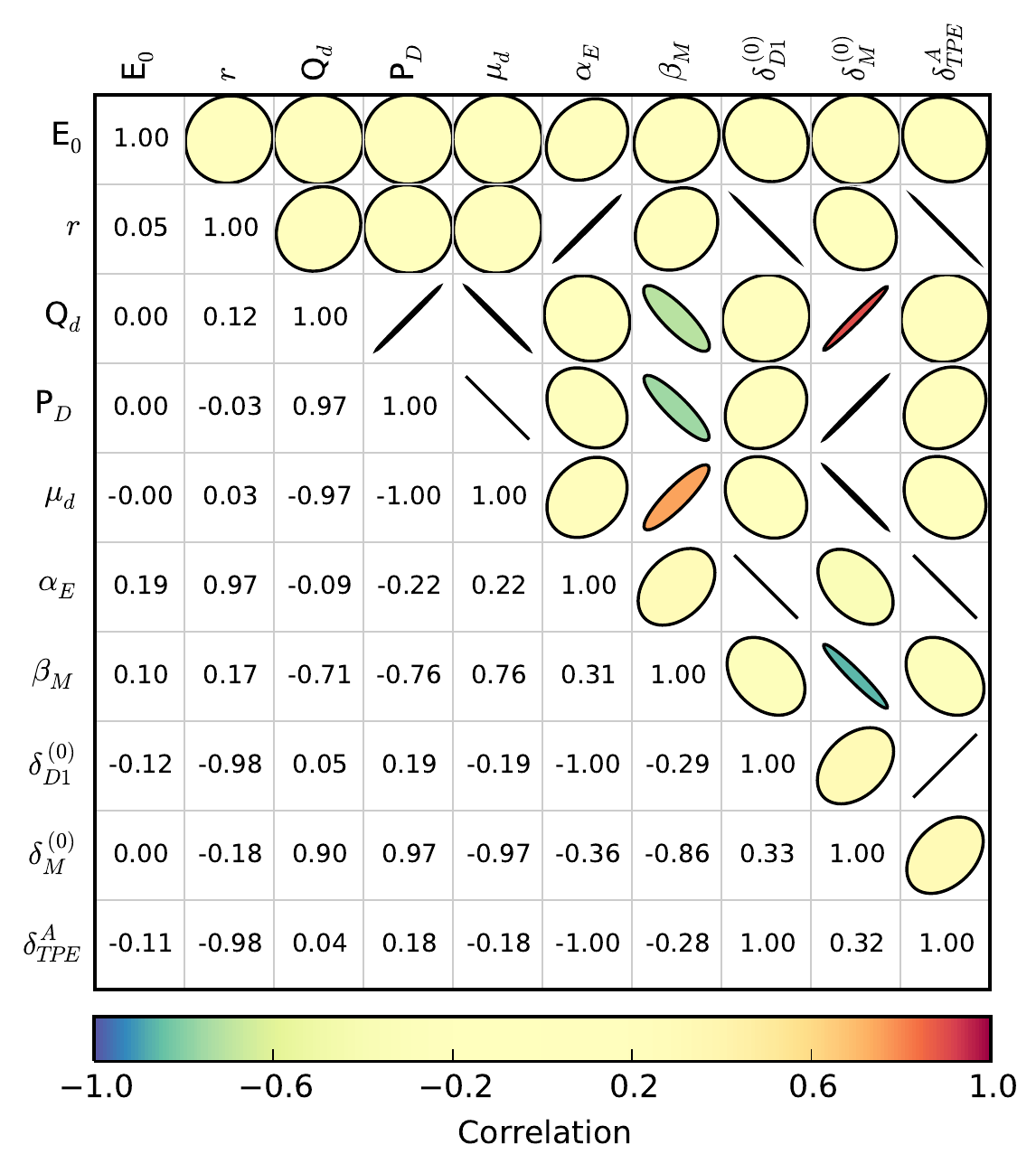}
      \caption{ The correlation matrix of the deuteron ground state energy
        $E_0$, rms radius $r$, quadrupole moment $Q_d$, D-state
        probability $P_D$, magnetic moment $\mu_d$,
      electric polarizability $\alpha_E$, magnetic susceptibility
      $\beta_M$, leading dipole polarizability correction
      $\delta^{(0)}_{D1}$ (by far the largest term in $\delta^{(0)}$, see
      Ref.~\cite{Hernandez_2014}), magnetic polarization correction
      $\delta^{(0)}_M$ and $\delta^{A}_{\rm TPE}$ for the $\NNLO_{sim}$
      potential with $\Lambda = 450$ MeV and $T^{max}_{lab}=125$ MeV.}
\label{fig:the N2LO systematic and statistical uncertainty estimates}
\end{figure}

We observe that the quadrupole moment $Q_d$ of the deuteron is strongly correlated with the $P_D$~\cite{Wong_1998}. 
The magnetic moment of the deuteron ground state $\mu_d$ can be
calculated analytically and depends only on the $P_D$ probability
\cite{Wong_1998}. As expected, from our numerical analysis, the
correlation $\rho(\mu_d,P_D)$ is strongly negative. Furthermore, from this correlation 
we also expect to see a correlation between $P_D$ and
$\delta^{(0)}_M$ which we indeed observe. Based on this
analysis, we see that the magnetic properties of the deuteron are
strongly related to $P_D$, indicating that they are largely
determined by the $D$-wave component of the deuteron. By contrast, the electric properties are found to be related to $r$. For example, a correlation is observed between $r$ and $\alpha_E$. This correlation is predicted on the basis of the zero-range model of the deuteron
\cite{Martorell_1995} and has been observed to hold in heavier
systems~\cite{Hagen2016,Miorelli_2016}. In fact, we find that $\delta^A_{\rm TPE}$,
which is strongly dominated by the dipole term $\delta^{(0)}_{D1}$, is correlated to $r$ and $\alpha_E$.

At this point the production of expected correlations within the
formalism of the statistical uncertainty propagation served as a way
to inspect the validity of our statistical analysis, but it could also
be of guidance if one decided to use alternative fitting procedures in
the future.

The statistical uncertainties of $r$, the electric dipole
polarizability $\alpha_E$, and $\delta^A_{\rm TPE}$ are $0.02\%$,
$0.05\%$, and $0.05\%$ respectively, i.e., negligible compared to the
size of the systematic uncertainty, as we will show in the next
Section. Importantly, if one uses the separately (or
  sequentially) optimized N$^2$LO$_{sep}$ potentials of
  Ref.~\cite{Carlsson_2016}, which fit the LECs in the $\pi$N, NN and
  3N sectors separately, the deduced statistical errors would be
  larger. This originates from neglecting the statistical covariances
  between the $\pi$N and NN (and 3N) sectors of the chiral EFT
  potentials while also employing LECs with rather large statistical
  uncertainties. Most microscopic interactions are constrained to data
  in such a sequential manner, in particular the chiral EFT potentials
  up to 5$^{\rm th}$ order that we employ in the next
  Section. However, the sub-leading $\pi$N LECs employed in
  those potentials were separately optimized using a novel Roy-Steiner
  extrapolation of the $\pi$N scattering
  data~\cite{Hoferichter_2016}. The resulting $\pi$N LECs exhibit very
  small statistical uncertainties~\cite{Hoferichter_2015}, and a
  forward error propagation to $\delta^{A}_{\rm TPE}$ would most
  likely lead to similarly reduced statistical uncertainties.

\section{Systematic uncertainty estimates} 
\label{syst}
Here we present the systematic uncertainties in our calculations due
to various truncations introduced in chiral EFT. First, we address the
truncation via $T^{\text{max}}_{\text{Lab}}$ in the energy range of
the NN scattering data used in the fit of the LECs.
\begin{figure}[h]
\begin{center}
 \includegraphics[width=0.48\textwidth]{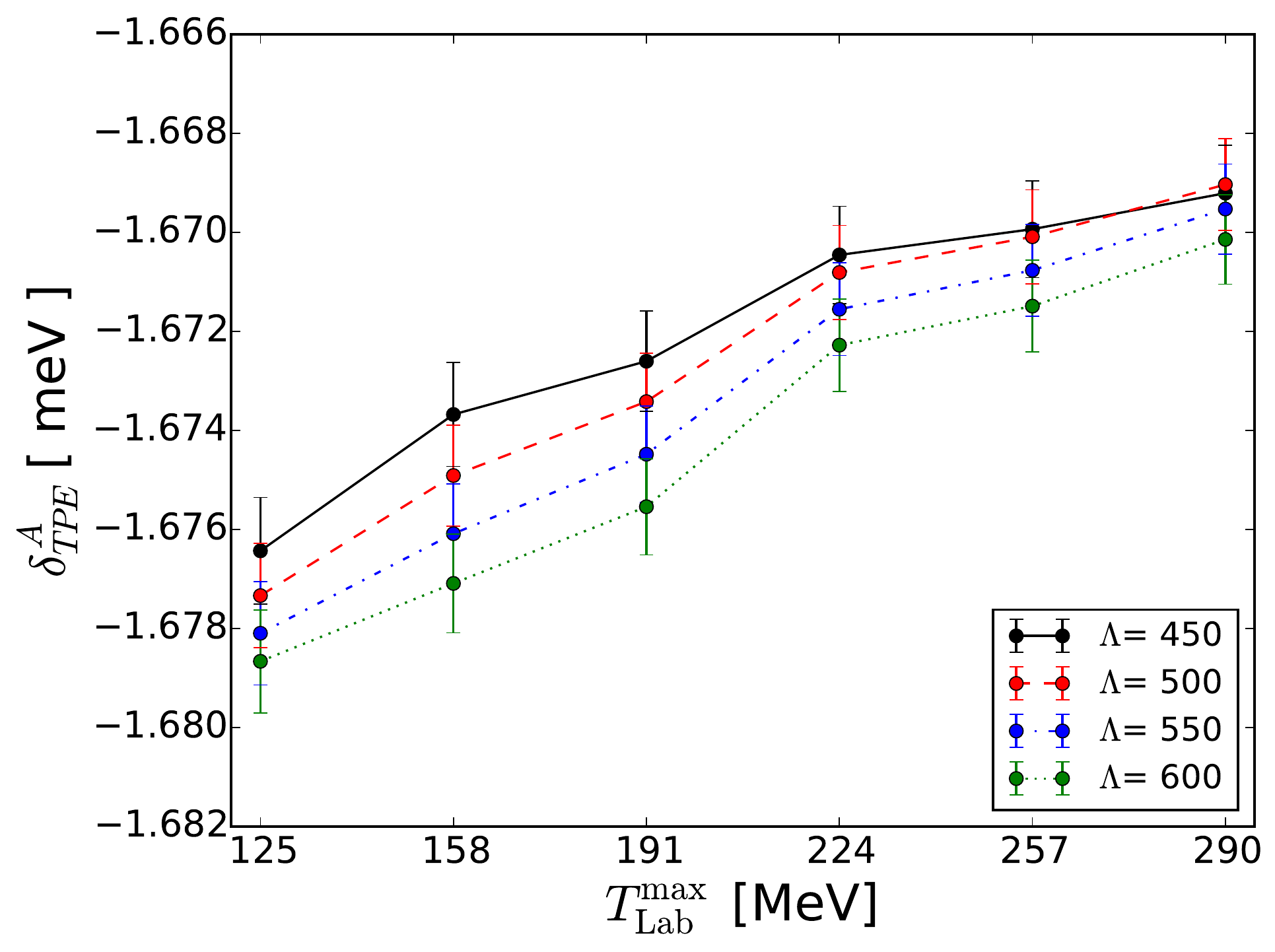}
\caption{The calculated values of $\delta^{A}_{\rm TPE}$ for different cutoffs $\Lambda$ in MeV as a function of $T^{\text{max}}_{\text{Lab}}$ for the N$^2$LO$_{sim}$ potentials. }
\label{fig:N2LOsim Nuclear Structure Uncertainties vs Tmax}
\end{center}
\end{figure}
In Fig.~\ref{fig:N2LOsim Nuclear Structure Uncertainties vs Tmax}, the
calculated values of $\delta^{A}_{\rm TPE}$ are plotted as a function
of the maximum lab energy, $T^{\text{max}}_{\text{Lab}}$, in the fit
of the N$^2$LO$_{sim}$ potentials for various choices of the cutoff
$\Lambda$. The error bars indicate the statistical uncertainties,
computed as detailed in the previous Section, which were on average
found to be 0.001 meV, or 0.06$\%$. In Fig.~\ref{fig:N2LOsim Nuclear Structure
  Uncertainties vs Tmax} it is clear that the statistical
uncertainties are small in comparison to the systematic uncertainties
due the variation of the cutoff $\Lambda$ and
$T^{\text{max}}_{\text{Lab}}$.
%ANDREAS: suggested to remove this paragraph. Not very clear, and is it necessary?
%Furthermore, the relative percent difference for 
%$\delta^{A}_{\rm TPE}$ at $T^{\text{max}}_{\text{Lab}}=125$ MeV and
%$T^{\text{max}}_{\text{Lab}}=290$ MeV for different cutoffs is
%0.5$\%$. This indicates that the bulk of necessary the physics for
%describing $\delta^{A}_{\rm TPE}$ is at least included in the fits of
%the NN-potential in the $125-290$ MeV energy range.
Furthermore, the range of the calculated values of $\delta^{A}_{\rm
  TPE}$ for different $\Lambda$ decreases at the largest
$T^{\text{max}}_{\text{Lab}}$ energies, indicating that the nuclear
dynamics as described by the LECs become better constrained with more
data.

Next, we address the uncertainties coming from truncating chiral EFT
at the order $\nu$. The common approach to gauge this uncertainty is
by varying the cutoff $\Lambda$ over a range of values. However, this
approach to uncertainty estimation suffers from several deficiencies,
such as the arbitrariness in the chosen $\Lambda-$range. Furthermore,
often the residual $\Lambda$ dependence underestimates uncertainties
as discussed in Refs.~\cite{Epelbaum_2015_01,Epelbaum_2015_02}. To
address these deficiencies and give a conservative estimate of the
systematic uncertainties, in addition to cutoff variation, we
follow~\cite{Furnstahl_2015,Epelbaum_2015_02} and include an
uncertainty estimate based on the expected size of the next
higher-order contribution in the chiral EFT expansion.
This approach is in semi-quantitative agreement with a Bayesian uncertainty analysis. 
Assuming that an observable $A(p)$, associated with an
external momentum scale $p$, and computed non-perturbatively from
chiral EFT, follows the same order-by-order pattern as chiral EFT
itself, then it can be expressed as
\begin{equation}
A(p) = A_0 \sum_{\nu=0}^{\infty} c_{\nu}(p) Q^{\nu},
\end{equation} 
where $A_{0}$ is the leading order value, $Q$ is the small expansion
parameter, given in Eq.~\eqref{eq: ChiEFT expansion parameter}, and
$c_{\nu}(p)$ is an observable- and interaction-specific expansion
coefficient determined  a posteriori. The uncertainty in $A$ due to
truncation at some finite order $\nu$, i.e., LO, NLO,
N$^{2}$LO,$\ldots$, can be estimated  by
\begin{equation}
\label{eq: Chiral Expansion Truncation uncertainty Estimate}
\sigma^{\text{N}^{k}\text{LO}}_{A,sys}(p) = A_{0} \cdot Q^{k+2} \text{max} \lbrace |c_0|,..., |c_{k+1}| \rbrace.
\end{equation}
This expression rests on a prior assumption of independent expansion coefficients $c_{\nu}$ with a boundless and uniform distribution.
%More explicitly, this leads to the following prescription at each order \cite{Epelbaum_2015_02}
%\begin{align}
%\sigma^{\NNNNLO}_{A,sys}(p) &= \text{max}\lbrace \ Q^{6}| A^{\LO}(p)|,  Q^{4}| A^{\LO}(p)-A^{\NLO}(p)|,\\
%\nonumber & Q^{3}| A^{\NLO}(p)-A^{\NNLO}(p)|, Q^{2}| A^{\NNLO}(p)-A^{\NNNLO}(p)|\\
%\nonumber & Q| A^{\NNNLO}(p)-A^{\NNNNLO}(p)| \ \rbrace \\
%\nonumber \sigma^{\NNNLO}_{A,sys}(p) &= \text{max}\lbrace \ Q^{5}| A^{\LO}(p)|,  Q^{3}| A^{\LO}(p)-A^{\NLO}(p)|,\\
%\nonumber & Q^{2}| A^{\NLO}(p)- A^{\NNLO}(p)|,Q| A^{\NNLO}(p)-A^{\NNNLO}(p)| \ \rbrace  \\
%\nonumber \sigma^{\NNLO}_{A,sys} (p) &= \text{max}\lbrace \ Q^{4}| A^{\LO}(p)|,  Q^{2}| A^{\LO}(p)-A^{\NLO}(p)|,\\
%\nonumber & Q| A^{\NLO}(p)-A^{\NNLO}(p)| \  \rbrace  \\
%\nonumber \sigma^{\NLO}_{A,sys}(p) &= \text{max}\lbrace \ Q^{3}| A^{\LO}(p)|,  Q| A^{\LO}(p)-A^{\NLO}(p)| \ \rbrace .
%\end{align}

To estimate the typical momentum scale $p$ of the nuclear structure
corrections, we compute the average energy value of the largest term
in $\delta^A_{\rm TPE}$, namely
the leading order dipole correction~\cite{Hernandez_2014}
\begin{equation}
\delta^{(0)}_{D1} = -\frac{2\pi m^3_r (Z \alpha)^5}{9}\int\limits_{\omega_{th}}^{\infty} d\omega \ \sqrt{\frac{2m_r}{\omega_N}} S_{D1}(\omega),
\end{equation}
where $S_{D1}(\omega)$ is the dipole response function. The average value $\langle \omega\rangle_{D1}$ is calculated as
\begin{equation}
\langle \omega \rangle_{D1} = \frac{\int d\omega \ \omega  \sqrt{\frac{2m_r}{\omega_N}}~ S_{D1}(\omega)}{\int d\omega \ \sqrt{\frac{2m_r}{\omega_N}}~ S_{D1}(\omega)}.
\end{equation}
Given that we obtain $\langle \omega \rangle_{D1} \approx 7 $ MeV, which corresponds to a momentum scale $p$ smaller than $m_\pi$, the
chiral convergence parameter $Q$ for our uncertainty estimates is
always taken to be $m_{\pi}/\Lambda_b$.
For a solid estimation of truncation errors, it is also crucial to adopt a suitable $\Lambda_b$. Here we follow \cite{Epelbaum_2015_02,Furnstahl_2015}  and set $\Lambda_b$ to 600 MeV, as a choice shown by Bayesian analyses \cite{Furnstahl_2015,Melendez_2017} to be optimal.

\begin{figure}[h]
\begin{center}
 \includegraphics[width=0.47\textwidth]{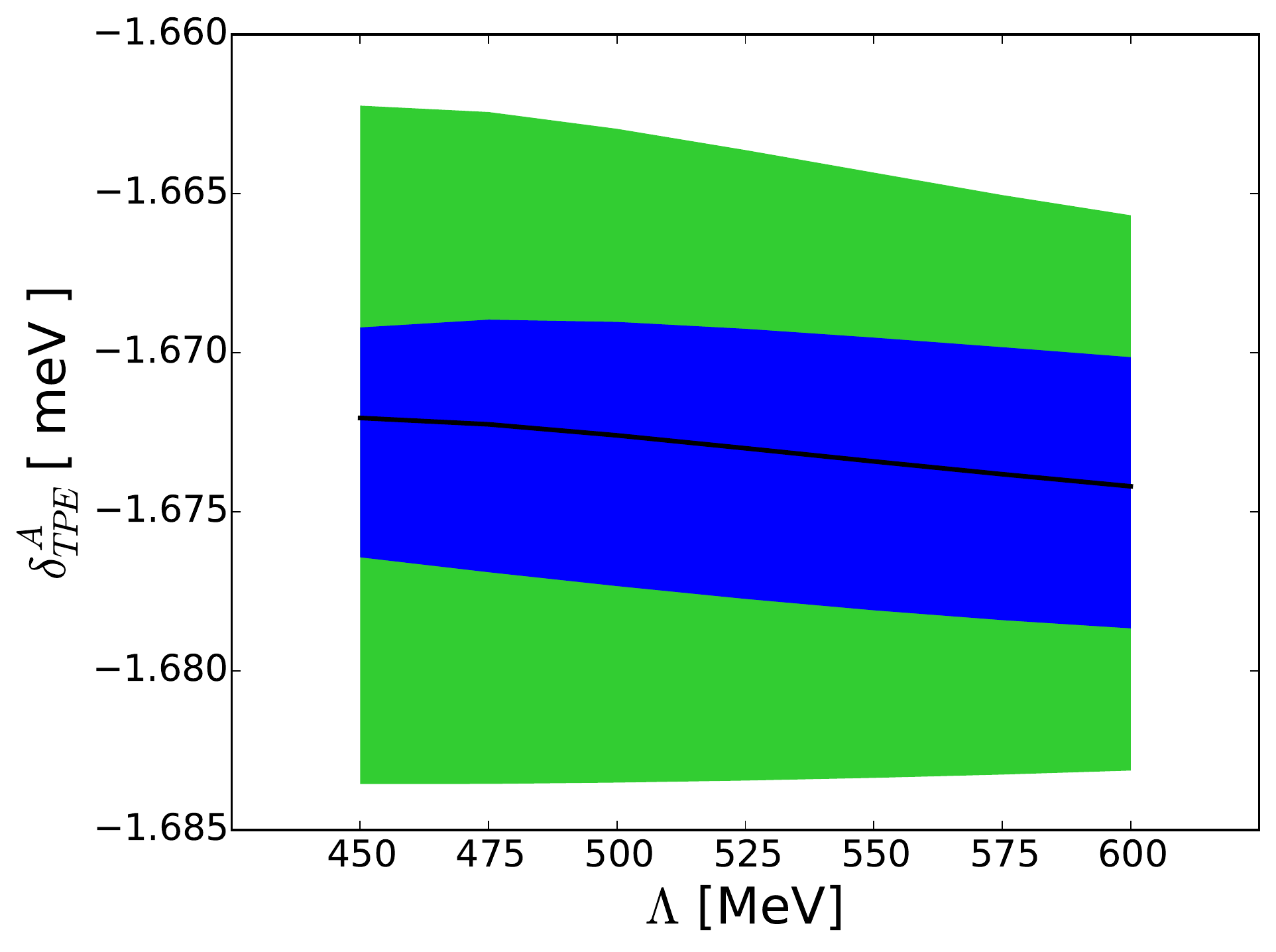}
\caption{Systematic uncertainties of $\delta^{A}_{\rm TPE}$ as a function of the cutoff $\Lambda$
  for the N$^2$LO$_{sim}$ potentials. The blue (dark) band indicates the 
  uncertainty due to
  variations in $T^{\text{max}}_{\text{Lab}}$. The (light) green band also includes the chiral truncation
  uncertainty.}
\label{fig:N2LOsim Nuclear Structure Uncertainties}
\end{center}
\end{figure}

In Fig.~\ref{fig:N2LOsim Nuclear Structure Uncertainties}, the
uncertainty estimates for $\delta^{A}_{\rm TPE}$ are displayed for the
N$^2$LO$_{sim}$ potentials as a function of the cutoff $\Lambda$. The
blue band shows a conservative spread of $\delta^A_{\rm TPE}$ due to
variations of $T^{\text{max}}_{\text{Lab}}$ from 125 to 290 MeV,
which, with respect to the central value, amounts to about $0.004$ meV
or 0.2$\%$.  The green band also includes the uncertainty stemming
from the chiral truncation. Due to the fact that the two systematic
uncertainties are not independent from each other, the chiral
truncation error is calculated from Eq.~\eqref{eq: Chiral Expansion
  Truncation uncertainty Estimate} and added to each point in
$T^{\text{max}}_{\text{Lab}}$ and $\Lambda$. The green band
encompasses the maximum and minimum values of $\delta^A_{\rm TPE}$ so
obtained.  The largest contribution of the truncation uncertainty
alone is found to be 0.007 meV for the N$^2$LO$_{sim}$ potentials.
%(similar values are also obtained for the other potentials at this order)
The overall systematic uncertainty
including cutoff variation, $T^{\text{max}}_{\text{Lab}}$ variations and chiral truncation error amounts to 0.011 meV or 0.65$\%$  for the  N$^2$LO$_{sim}$ potentials and  thus dominates with respect to the 0.06$\%$ statistical uncertainty.

To study the convergence of  $\delta^A_{\text{TPE}}$ with respect to
the chiral orders greater than N$^2$LO, in addition to the $sim$
potentials, we also carry out calculations using the chiral potentials
available up to N$^4$LO. Two groups of chiral interactions have been
constructed using different fitting procedures and slightly different
operatorial form in the potentials, but identical power-counting.  We
will use all orders available from Ref.~\cite{Epelbaum_2015_01}\footnote{This is a newer version of the potentials with respect to those we used in Ref.~\cite{Hernandez_2014}.} and
denote them as N$^k$LO$_{EKM}$, and those from Ref.~\cite{Entem_2017},
which will be denoted as N$^k$LO$_{EMN}$. For the  N$^k$LO$_{EKM}$ family of potentials we will explore the cutoffs $(R_0,\Lambda)=(0.8,600), (1.0,600)$ and $(1.2,400)$ [fm, MeV], where $R_0$ is a coordinate-space regulator, and for the N$^k$LO$_{EMN}$ family of potentials we will use $\Lambda=450,500$ and 550 MeV.
The use of a higher order in
chiral EFT will allow us not only to update our results with respect
to our previous work~\cite{Hernandez_2014,Hernandez_2016}, but also to
get a more reliable estimate the chiral convergence uncertainty using
Eq.~(\ref{eq: Chiral Expansion Truncation uncertainty
  Estimate}). Our goal is to provide an updated value of
$\delta_{\rm TPE}$ with its overall uncertainty. This will be
discussed in the next Section.

\section{Total uncertainty estimates}
\label{results}

First, systematic and statistical uncertainties of the various nuclear interactions are combined into $\sigma_{\rm Nucl}$, which is detailed in Table~\ref{table:Main Results of the paper}. 
For  the $\EKM$ and $\EMN$
potentials at all orders we include  systematic uncertainties
from chiral convergence and cutoff variation.  Systematic errors stemming from
$T^{\text{max}}_{\text{Lab}}$ variations cannot presently be estimated
at N$^3$LO and at N$^4$LO, thus we include the corresponding uncertainty
evaluated from N$^2$LO$_{sim}$.  For the lowest orders  in the $EKM$ and $EMN$ potentials, systematic errors from
$T^{\text{max}}_{\text{Lab}}$ variations are taken from the corresponding order of the $sim$ interactions.
The $sim$
potentials contain all of the above and statistical uncertainties, estimated consistently at each order.
Statistical uncertainties are found to be negligible in the $sim$ potentials and, while at present a consistent evaluation is not possible, they are also expected to be small in the  $\EKM$ and $\EMN$ families. Thus, we take the  N$^2$LO$_{sim}$ statistical values also for the  N$^{2,3,4}$LO$_{EKM/ EMN}$ potentials, and the LO$_{sim}$ (NLO$_{sim}$) values for the LO$_{EKM/ EMN}$ (NLO$_{EKM/ EMN}$) potentials, respectively.
%%%%%
\begin{table}[h]
\caption{Results for $\delta_{\rm TPE}$ at various orders with corresponding estimates for the nuclear physics $\sigma_{\rm Nucl}$ and total $\sigma_{\text{Tot}}$  uncertainties.}
\label{table:Main Results of the paper}
\renewcommand{\tabcolsep}{0.7mm}
\footnotesize
\begin{center}
\begin{tabular}{l c c c c}
\hline\hline
Order  & Potential  & $\delta_{\text{TPE}}$ & $\sigma_{\text{Nucl}}$    &  $\sigma_{\text{Tot}}$  \\ 
  &   &  [meV]  & [meV] & [meV]   \\ \hline 
$\LO$        & $\begin{matrix}
 {sim} \\[3pt] 
 {EKM} \\[3pt] 
 {EMN}
\end{matrix}$ & $\begin{matrix}
 \LOsimAvg \\[3pt]
 \LOekmAvg \\[3pt] 
 \LOemnAvg 
\end{matrix}$ &  $\begin{matrix}
  _{\LOsimMin}^{\LOsimMax}\\[3pt]
  _{\LOekmMin}^{\LOekmMax}\\[3pt]
  _{\LOemnMin}^{\LOemnMax}
\end{matrix}$ & $\begin{matrix}
  _{\LOsimMinTot}^{\LOsimMaxTot}\\[3pt]
  _{\LOekmMinTot}^{\LOekmMaxTot}\\[3pt]
  _{\LOemnMinTot}^{\LOemnMaxTot}
\end{matrix}$ \Tstrut\Bstrut\\ \hline
$\NLO$ & $\begin{matrix}
 {sim} \\[3pt]
 {EKM} \\[3pt]
 {EMN}
\end{matrix}$ & 
$\begin{matrix}
 \NLOsimAvg \\[3pt]
 \NLOekmAvg \\[3pt]
 \NLOemnAvg
\end{matrix}$
 & $\begin{matrix}
  _{\NLOsimMin}^{\NLOsimMax}\\[3pt]
  _{\NLOekmMin}^{\NLOekmMax}\\[3pt]
  _{\NLOemnMin}^{\NLOemnMax}
\end{matrix}$ &
$\begin{matrix}
  _{\NLOsimMinTot}^{\NLOsimMaxTot}\\[3pt]
  _{\NLOekmMinTot}^{\NLOekmMaxTot}\\[3pt]
  _{\NLOemnMinTot}^{\NLOemnMaxTot}
\end{matrix}$ \Tstrut\Bstrut\\ \hline
 $\NNLO$ & $\begin{matrix}
 {sim} \\[3pt]
 {EKM} \\[3pt]
 {EMN}
\end{matrix}$ & 
$\begin{matrix}
 \NNLOsimAvg \\[3pt]
 \NNLOekmAvg \\[3pt]
 \NNLOemnAvg
\end{matrix}$
&  
 $\begin{matrix}
  _{\NNLOsimMin}^{\NNLOsimMax}\\[3pt]
  _{\NNLOekmMin}^{\NNLOekmMax}\\[3pt]
  _{\NNLOemnMin}^{\NNLOemnMax}
\end{matrix}$ &
$\begin{matrix}
  _{\NNLOsimMinTot}^{\NNLOsimMaxTot}\\[3pt]
  _{\NNLOekmMinTot}^{\NNLOekmMaxTot}\\[3pt]
  _{\NNLOemnMinTot}^{\NNLOemnMaxTot}
\end{matrix}$
\Tstrut\Bstrut\\ \hline
  $\NNNLO$ & 
  $\begin{matrix}
 {EKM} \\[3pt]
 {EMN}
\end{matrix}$ & 
$\begin{matrix}
 \NNNLOekmAvg \\[3pt]
 \NNNLOemnAvg 
\end{matrix}$
 &    $\begin{matrix}
  _{\NNNLOekmMin}^{\NNNLOekmMax} \\[3pt]
  _{\NNNLOemnMin}^{\NNNLOemnMax} 
\end{matrix}$ &
$\begin{matrix}
  _{\NNNLOekmMinTot}^{\NNNLOekmMaxTot}\\[3pt]
  _{\NNNLOemnMinTot}^{\NNNLOemnMaxTot}
\end{matrix}$
\Tstrut\Bstrut\\ \hline
   $\NNNNLO$ & $\begin{matrix}
 {EKM} \\[3pt]
 {EMN}
\end{matrix}$ &
$\begin{matrix}
 \NNNNLOekmAvg \\[3pt]
 \NNNNLOemnAvg 
\end{matrix}$
 &  
 $\begin{matrix}
  _{\NNNNLOekmMin}^{\NNNNLOekmMax} \\[3pt]
  _{\NNNNLOemnMin}^{\NNNNLOemnMax} 
\end{matrix}$ &
$\begin{matrix}
  _{\NNNNLOekmMinTot}^{\NNNNLOekmMaxTot}\\[3pt]
  _{\NNNNLOemnMinTot}^{\NNNNLOemnMaxTot}
\end{matrix}$
\Tstrut\Bstrut\\ \hline \hline
%$\NNNNLO$ & 
%$ \text{Total}$ &
%$ \FinalAvg$ &
%$ _{\FinalMin}^{\FinalMax}$ &
%$ _{\FinalMinTot}^{\FinalMaxTot}$\\
%\hline \hline
\end{tabular}
\end{center}
\end{table}

The present calculation is performed to sub-subleading order in
the $\eta$-expansion, thus uncertainties deriving from higher order $\eta$ contributions need to be
estimated in the total error budget. These higher
order $\eta$ contributions are estimated to provide a $0.3\%$ effect based on a different approach to the computation of
$\delta^A_{\rm TPE}$, which allows to include higher order
electromagnetic multipoles~\cite{Javier_inprep}.
So far  we have
concentrated on $\delta^A_{\rm TPE}$, which is the only term in
$\delta_{\rm TPE}$ with explicit dependence on the nuclear
dynamics. For a complete discussion on $\delta_{\rm TPE}$  we should consider the additional nucleonic terms in
Eq.~(\ref{TPE_N}), namely $\delta^{N}_{\text{pol}}$, $\delta^{N}_{Zem}$,
$\delta^{N}_{\text{sub}}$ and their respective uncertainties, using the values quoted in
Section~\ref{tpe}.
Finally, our $\delta_{\rm TPE}$ formulas are valid in an $\alpha$ expansion up to
5$^{th}$ order. Higher order terms in the $\alpha$ expansion were
estimated first by Pachucki~\cite{Pachucki_2011} to be of 1$\%$. Here,
we will keep this value and refer to it as the atomic physics
uncertainty, as in Refs.~\cite{Hernandez_2014,Hernandez_2016}, and add
it to the other uncertainties in quadrature.
Atomic,  single-nucleon, and $\eta$-expansion uncertainties
 of $\delta_{\rm TPE}$ are included in $\sigma_{\rm Tot}$, on top of the nuclear physics uncertainties, given in Table~\ref{table:Main Results of the paper}.

\begin{figure}[h]
  \centering
    \includegraphics[scale=0.48]{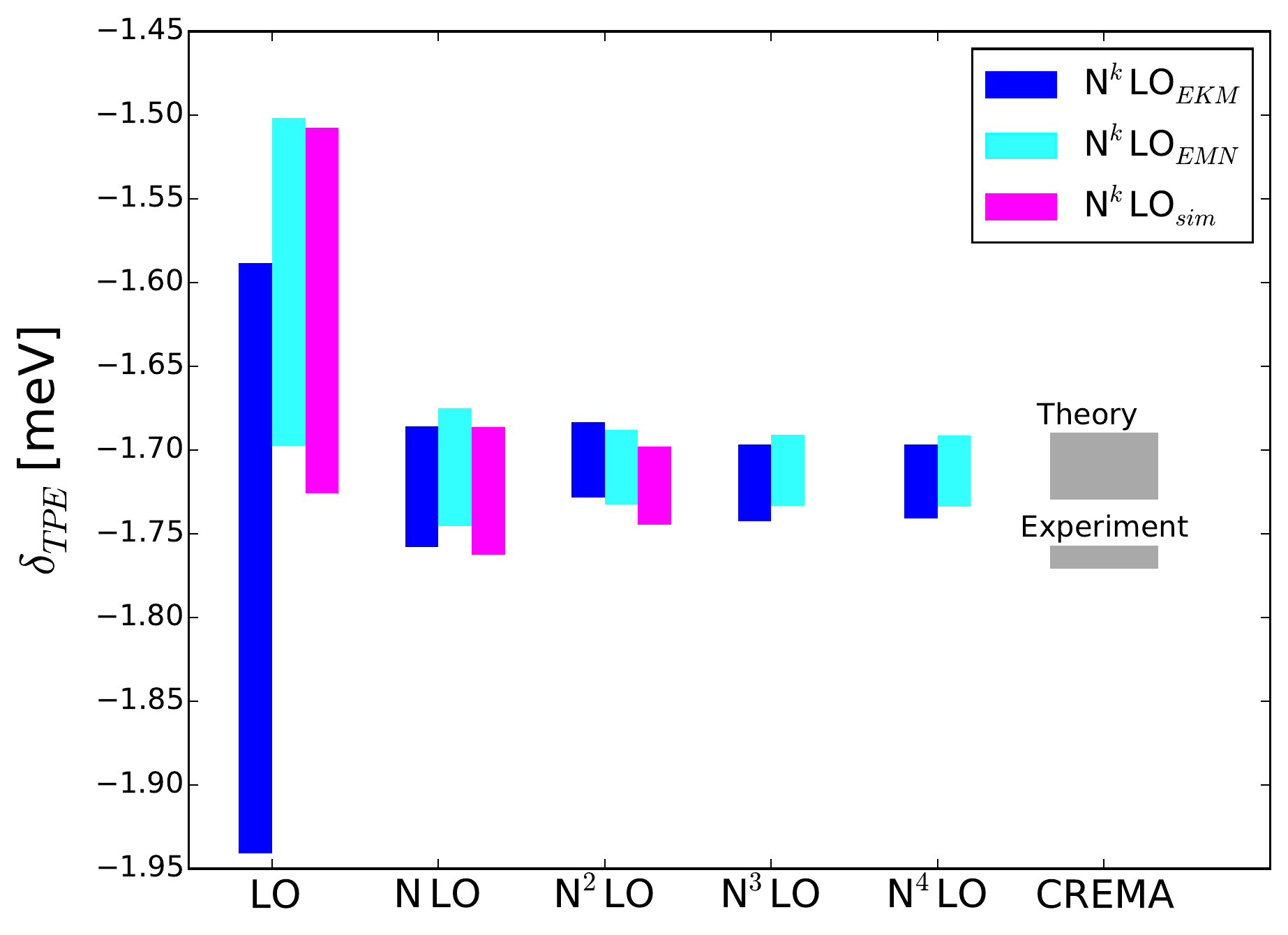}
      \caption{ $\delta_{\text{TPE}}$ as a function of the chiral
        order with total uncertainty (see text for details).}
      \label{fig: The convergence of the final delta TPE}
\end{figure}

In Fig.~\ref{fig: The convergence of the final delta TPE} and in Table~\ref{table:Main Results of the
  paper}, we show the convergence of $\delta_{\text{TPE}}$ and its overall uncertainties
with respect to all chiral orders from LO up to N$^4$LO using $\delta^N_{sub}$ from Ref.~\cite{Krauth_2016} (we will include the larger uncertainties of Ref.~\cite{Hill:2016bjv} later in our analysis). In particular, in Table \ref{table:Main Results of the
  paper}, one can appreciate the difference between $\sigma_{\rm Nucl}$ and $\sigma_{\rm Tot}$.
One can readily see that uncertainty bands decrease as the order of
the chiral expansion increases, as expected.  Beside LO calculations,
where the three potential families somehow differ, all results at higher
orders are quite stable around the same value, independently of the
potential used. Interestingly, N$^4$LO results are almost identical to
N$^3$LO results, indicating convergence of the chiral expansion for
this observable. The uncertainty estimates at N$^3$LO and N$^4$LO are
compatible with our previous estimates in Refs.~\cite{Hernandez_2014,Hernandez_2016}, even though slightly larger,
mostly due to the inclusion of the systematic error using
Eq.~(\ref{eq: Chiral Expansion Truncation uncertainty Estimate}).
Furthermore, Table~\ref{table:Main Results of the paper}  shows that, although the nuclear physics errors 
are dominant at lower order in the chiral expansion, at
N$^4$LO the leading uncertainty is not stemming from
nuclear physics, but rather from the other sources.

Results are also compared to the experimentally inferred $\delta_{\rm
  TPE}$ correction \cite{Pohl_2016} and theoretical compilation
\cite{Krauth_2016}.  We find that the N$^4$LO band is consistent with
the theoretical compilation and encompasses also our result
$\delta_{\rm TPE}=-1.709$ meV from Ref.~\cite{Hernandez_2014} based on the AV18
potential~\cite{AV18}, which is also included in the theory summary by
Krauth~\cite{Krauth_2016}. We also observe, though, that our N$^4$LO
band is not compatible with the experimental determination of
$\delta_{\rm TPE}$.

\begin{table}[htb]
  \caption{Uncertainty breakdown of the final  $\delta_{\text{TPE}}$ value. For the single-nucleon contribution we quote two values, one where we adopted the strategy of Ref.~\cite{Krauth_2016} and one where we use the larger uncertainties from Ref.~\cite{Hill:2016bjv} for $\delta^N_{sub}$.}
\label{table:The final uncertainty budget}
\renewcommand{\tabcolsep}{0.7mm}
\footnotesize
\begin{center}
\begin{tabular}{l c l l}
\hline\hline
Contribution &  & Uncertainty in meV & \\
\hline 
Nuclear physics (syst) &   & 
$\begin{matrix}
  +0.008 \\[0.5pt]
-0.011
\end{matrix}$ & \\
\hline
{Nuclear physics (stat)} & & {$\pm$0.001} &\\
\hline
$\eta$-expansion & & {$\pm$0.005} &\\
\hline
Single-nucleon & & {$\pm$0.0102}~\cite{Krauth_2016} & {$\pm$0.0198}~\cite{Hill:2016bjv} \\
Atomic physics    & & $\pm$0.0172 & \\
\hline 
Total  &  & $\begin{matrix} 
  +0.022 \\[0.5pt]
  -0.024 
\end{matrix}$ &
 $\begin{matrix} 
  +0.028 \\[0.5pt]
  -0.029 
\end{matrix}$
\\
\hline \hline
\end{tabular}
\end{center}
\end{table}

Finally, based on our analysis, we provide an updated value of
$\delta_{\text{TPE}}=-1.715$ meV with its itemized uncertainty budget in Table
\ref{table:The final uncertainty budget}.  As central value we take
the average N$^4$LO result from the $EMN$ and $EKM$
families. Uncertainties are separated into systematic and statistic
nuclear physics, $\eta$-expansion, single-nucleon and atomic physics
uncertainties.
Systematic uncertainties from cutoff variation and chiral truncation
are obtained from our N$^4$LO studies by taking the combined range of
the $EMN$ and $EKM$ bands.  While here we studied the chiral convergence of
the potential, the same should in principle be done regarding the
chiral expansion of electromagnetic
currents~\cite{Pastore08,Kolling11} leading to further
systematic corrections.  In our formalism, $\delta^A_{\rm TPE}$ is
related to electromagnetic multipoles, where the electric dipole
dominates. The latter is protected by the Siegert
theorem~\cite{Arenhovel_book}, so that two-body currents are
implicitly included via the use of the continuity
equation~\cite{Arenhovel_book}. There exists corrections to the
Siegert term.  We have estimated the magnitude of those by integrating
an E1-response function provided by
Arenh{\"o}vel~\cite{Arenhovel_private}, which included two-body
currents and relativistic corrections as in Ref.~\cite{Arenhovel_book}
for the AV18 potential. Their effect on the leading dipole correction
$\delta^{(0)}_{D1}$ was both found to be of the order of 0.05$\%$,
thus negligible.

Despite the disputed single-nucleon TPE uncertainty~\cite{Hill:2016bjv,Birse:2017czd,Hill:2017rlj},  it is evident from Table 2 that the atomic physics error remains a major source of uncertainty.
It is approximately $1\%$ from a
reasonable, but rough, estimate by Pachucki {\it et
  al.}~\cite{Pachucki_2011}.
The estimate is based on taking 50$\%$ of
relativistic and higher order corrections, which are the smallest
contributions to $\delta^A_{\rm TPE}$.
A more thorough estimate 
of $\alpha^6$ effects 
requires going to third order in perturbation theory and study
three-photon exchange effects, which is beyond the scope of
this work. Here, we have shown that uncertainties stemming from the chiral
EFT description of the nuclear interaction alone are not capable of
explaining the discrepancy between the calculated $\delta_{\rm TPE}$ and
the corresponding experimental extraction by Pohl {\it et
  al.}~\cite{Pohl_2016}.

Since the deuteron point-nucleon radius $r$ based on CODATA was used in the fitting procedure, e.g., of $\NNLO_{sim}$ potentials, one may suspect this yields a biased $\delta_{\rm TPE}$ in muonic atoms. However, in order to remedy the discrepancy between the $\mu-d$ and ``$\mu p+$iso'' values of $r_d$, one may just  vary $r$  by  $\sim 0.1\%$, see Eq.~(\ref{eq_radius}). Due to the linear correlation between $r$ and $\delta_{\rm TPE}$, this would only lead to a maximum  variation $\delta_{\rm TPE}$ of the order of $\sim 0.1\%$, which is negligible with respect to the required $\sim 3\%$ change needed to explain the discrepancy between its theory and experimental determination.

\section{Conclusions}
\label{conclusion}

In this work, we have explored the uncertainties in 
$\delta_{\text{TPE}}$ corrections using state-of-the-art chiral
potentials from LO up to N$^4$LO. We have calculated the statistical
uncertainties up to N$^2$LO using a set of simultaneously optimized
chiral potentials. From this we conclude that the uncertainty due
to variances of the LECs are negligible compared to the systematic
uncertainties due to cutoff variation and chiral truncation. We have
also found that going beyond N$^{3}$LO in chiral EFT does not change
the overall results of $\delta_{\text{TPE}}$, which also indicates a
high theoretical accuracy of our final result. In conclusion, the rigorous
uncertainty quantification presented here weakens
the disagreement between  the calculated two-photon exchange
correction and the corresponding experimentally inferred value by Pohl
{\it et al.}~\cite{Pohl_2016} from 2.6 $\sigma$ to within 2
$\sigma$
(or 1.7~$\sigma$ if using the larger single nucleon uncertainties of Ref.~\cite{Hill:2016bjv}). Breaking down the total uncertainty budget in the calculation of $\delta_{\rm TPE}$ shows that atomic physics and single-nucleon physics need to be addressed to further reduce the theoretical uncertainty. It is important to remark that the deuteron-radius puzzle is still alive, in that the large discrepancy between the spectroscopic measurements on muonic deuterium and on ordinary deuterium still exists and it does not seem to be simply explained from nuclear physics uncertainties in the few-body dynamics.

\section{Acknowledgments}
\noindent
We would like to thank Angelo Calci for providing us with the chiral
potentials at N$^4$LO. We are grateful to Randolf Pohl and Hartmuth
Arenh\"{o}vel for useful discussions. This work was supported in parts
by the Natural Sciences and Engineering Research Council (NSERC), the
National Research Council of Canada, by the Deutsche
Forschungsgemeinschaft DFG through the Collaborative Research Center
[The Low-Energy Frontier of the Standard Model (SFB 1044)], and
through the Cluster of Excellence [Precision Physics, Fundamental
  Interactions and Structure of Matter (PRISMA)], by the Swedish
Research Council under Grant No. 2015- 00225, and by the Marie
Sklodowska Curie Actions, Cofund, Project INCA 600398.

%%\section*{References}

\bibliography{mybibfile}

\end{document}